\documentclass[a4paper,11pt]{article}
\usepackage{pos}
\usepackage{bbm} 
\newcommand{\tr}[1]{\left\langle #1 \right\rangle}
\newcommand{\com}[2]{\left[ #1,#2 \right]}
\newcommand{\acom}[2]{\left\{ #1,#2 \right\}}
\newcommand{\sdot}[2]{\left( #1\cdot #2 \right)}
\newcommand{\lt}[1]{\tilde{\lambda}_{#1}}

\renewcommand{\logo}{\relax}

\title{Axions in Baryon Chiral Perturbation Theory}

\author*[a]{Thomas Vonk}

\affiliation[a]{Helmholtz-Institut f\"{u}r Strahlen- und Kernphysik and Bethe Center for Theoretical Physics,
   Universit\"{a}t Bonn,\\Nussallee 14--16, D-53115 Bonn, Germany}

\emailAdd{vonk@hiskp.uni-bonn.de}

\abstract{I give an overview of recent developments in the study of the coupling of the QCD axion to nucleons and other octet baryons. I demonstrate how axions can be included into heavy baryon chiral perturbation theory, and present recent numerical results for the several axion-baryon couplings in SU(2) and SU(3) chiral perturbation theory for the Kim--Shifman--Vainstein--Zakharov (KSVZ) and the Dine--Fischler--Srednicki--Zhitnitsky (DFSZ) axion model.}

\FullConference{%
  RDP online PhD school and workshop ``Aspects of Symmetry'' (Regio2021),\\
  8-12 November 2021\\
  Online}

\begin{document}
\maketitle

\section{From the QCD \texorpdfstring{$\theta$}{Theta}-vacuum to the QCD axion}
Soon after the discovery of the instanton solution of non-Abelian gauge field theories \cite{Belavin:1975fg}, it became clear that the theory of the strong interaction, Quantum Chromodynamics (QCD), has a rich, complex vacuum structure \cite{Callan:1976je,Callan:1977gz}. While this instanton solution was the starting point for the resolution of the U(1)$_A$ problem proposed by t'Hooft \cite{tHooft:1976rip,tHooft:1976snw}, it at the same time lead to a new, still unresolved issue in QCD: the strong CP-problem. QCD with the $\theta$-term is not CP-invariant as a non-vanishing $\theta$ angle leads to a contribution to the neutron electric dipole moment (nEDM) $\propto \bar{\theta}=\theta + \operatorname{Arg}\operatorname{det}\mathcal{M}$ \cite{Baluni:1978rf} and affects meson and nucleon masses, as well as nuclear interactions and stellar and Big Bang nucleosynthesis \cite{Lee:2020tmi,Elhatisari:2021eyg}. However, theoretical estimations of the nEDM, which roughly vary between $|d_n| \approx 10^{-16} \bar{\theta}\,e\,\text{cm}$ and $|d_n| \approx 10^{-15} \bar{\theta}\,e\,\text{cm}$~\cite{Kim:2008hd,Guo:2015tla}, and
recent measurements yielding $|d_n^\text{exp}| < 1.8\times 10^{-26}\, e\,\text{cm}$~(90\,\% C.L.) \cite{nEDM:2020crw}, imply that $\bar{\theta} \lesssim 10^{-11}$. This is indeed a remarkable result and certainly in need of explanation: Why $\bar{\theta}$ is such a small quantity, whereas in principle it might take on any value between $[-\pi,\pi]$? It seems that there are at least no anthropic constraints that require $\theta$ to be of $\mathcal{O}(10^{-11})$ \cite{Ubaldi:2008nf,Lee:2020tmi}, not even $\mathcal{O}(10^{-2})$ (though $\bar{\theta}$ of $\mathcal{O}(1)$ \textit{has} considerable impact on the universe \cite{Lee:2020tmi}). Other alternatives, such as a vanishing quark mass seem to be improbable (see, e.g. \cite{Alexandrou:2020bkd}, and references therein).

The Peccei--Quinn (PQ) mechanism~\cite{Peccei:1977hh,Peccei:1977ur} coming with a new global chiral symmetry, now often labeled U(1)$_\text{PQ}$, is another solution to the strong-CP problem that is vividly discussed up to the present day. One of the reasons is that the pseudoscalar Nambu--Goldstone boson resulting from the spontaneous PQ-symmetry breakdown, called the axion~\cite{Weinberg:1977ma,Wilczek:1977pj}, became a reasonable dark matter candidate \cite{Preskill:1982cy,Abbott:1982af,Dine:1982ah,Ipser:1983mw,Turner:1986tb,Duffy:2009ig,Marsh:2015xka}. This in particular applies to the canonical ``invisible'' axion models, the KSVZ axion model~\cite{Kim:1979if,Shifman:1979if} and the DFSZ axion model~\cite{Dine:1981rt,Zhitnitsky:1980tq}, but also in general to any kind of axion-like particles (ALPs).

In QCD/QED, the traditional axions couple to gluons, quarks, and photons, which means that on a less fundamental level, axions also interact with mesons, nucleons, and other baryons. Here I discuss recent developments in the study of the axion-nucleon coupling \cite{Donnelly:1978ty,Kaplan:1985dv,Srednicki:1985xd,Georgi:1986df,GrillidiCortona:2015jxo} in SU(2) heavy baryon chiral perturbation theory (HBCHPT) and the general axion-baryon coupling in SU(3) HBCHPT as developed in Ref.~\cite{Vonk:2020zfh,Vonk:2021sit}. This is motivated from the fact that the axion-nucleon coupling plays a crucial role in determining the traditional {axion window} \cite{Preskill:1982cy,Abbott:1982af,Kim:1986ax}
\begin{equation}
10^{9}\,\text{GeV} \lesssim f_a \lesssim 10^{12}\,\text{GeV},
\end{equation}
where $f_a$ is the axion decay constant, whose large value causes the axion's weak coupling to Standard Model particles, but also the smallness of its mass. The process predominantly considered in the literature in the context of the axion-nucleon coupling is nuclear bremsstrahlung in massive stellar objects (see the overviews~\cite{Turner:1989vc,Raffelt:1990yz,Kim:2008hd,DiLuzio:2020wdo}), which requires a precise knowledge of the strength and structure of the axion-nucleon coupling. This is also true for axion-nucleon scattering 
\begin{equation}
aN\to \pi N
\end{equation}
which might be of some relevance in protosupernova cores, as has been proposed recently~\cite{Carenza:2020cis,Fischer:2021jfm}. Moreover, it has been suggested that there might be a considerable amount of hyperons existing in the cores of neutron stars (see, e.g., \cite{Tolos:2020aln} and references therein). Provided this is indeed the case, this would make it necessary to consider also the axion's coupling to baryons other than the proton and neutron when determining the axion's contribution to the cooling of neutron stars.

\section{QCD with axions}
After a suitable chiral rotation, the axion-quark interaction part of the QCD Lagrangian below the PQ symmetry breaking scale can be written in matrix notation as
 \begin{equation}
  \mathcal{L}_{aq} = - \left(\bar{q}_L\mathcal{M}_a q_R + \text{h.c.}\right) + \bar{q}\gamma^\mu
  \gamma_5 \frac{\partial_\mu a}{2f_a} \left(\mathcal{X}_q- \mathcal{Q}_a 
\right) q ,
\end{equation}
where $q=(u,d,s,c,b,t)^\mathrm{T}$ collects the quark spinors, $a$ is the axion field, and
\begin{equation}\label{eq:Ma}
\mathcal{M}_a =\exp\left({i\frac{a}{f_a}\mathcal{Q}_a}\right) \, \mathcal{M} ,\qquad
  \mathcal{Q}_a=\frac{\mathcal{M}^{-1}}{\tr{\mathcal{M}^{-1}}}\approx
  \frac{1}{1+z+w}\operatorname{diag}\left(1,z,w,0,0,0\right) .
\end{equation}
Here, $\mathcal{M} = \operatorname{diag}\left\{m_q\right\}$ is the $6\times 6$ quark mass matrix, and $z=m_u/m_d$ and $w=m_u/m_s$ (this particular form of $\mathcal{Q}_a$ has been chosen in order to avoid a mixing between the axion and the neutral Nambu--Goldstone bosons of chiral symmetry breaking). Furthermore, $\mathcal{X}_q = \operatorname{diag}\left\{X_q\right\}$ is the $6\times 6$ axion-quark coupling matrix. For the models under consideration, the KSVZ and the DFSZ, one has
 \begin{equation}
X_q^\mathrm{KSVZ} = 0, \qquad X_{u,c,t}^\mathrm{DFSZ} = \dfrac{1}{3}\sin^2\beta, \qquad
X_{d,s,b}^\mathrm{DFSZ}  = \dfrac{1}{3}\cos^2\beta = \dfrac{1}{3}
- X_{u,c,t}^\mathrm{DFSZ},
\end{equation}
where $\beta$ is related to the vacuum expectation values of the two Higgs doublets in the DFSZ model. 

From this Lagrangian one has to determine the external fields $a_\mu$ and $a_\mu^{(s)}$ that enter chiral perturbation theory. In SU(2) this can be achieved by separating the 2-dimensional flavor subspace of the two lightest quarks from the rest and by decomposing the matrix $\left(\mathcal{X}_q- \mathcal{Q}_a \right)$ into traceless parts and parts with non-vanishing trace, which results in
\begin{align}
\begin{split}
  \mathcal{L}_{aq} =  - \left(\bar{q}_L\mathcal{M}_a q_R  + \text{h.c.}\right) 
 & + \left( \bar{q}\gamma^\mu\gamma_5  \left( c_{u-d} \frac{\partial_\mu a}{2f_a} \tau_3 + c_{u+d} \frac{\partial_\mu a}{2f_a} \mathbbm{1}\right)
   q\right)_{q=(u,d)^\mathrm{T}} \\ &+ \sum_{q=\{s,c,b,t\}} \left(  \bar{q}\gamma^\mu \gamma_5 c_q  \frac{\partial_\mu a}{2f_a} q\right)  
\end{split}
\end{align}
with $\tau_3$ being the third Pauli matrix and
\begin{equation}
c_{u\pm d} = \frac{1}{2}\left(X_u \pm X_d-\frac{1\pm z}{1+z+w}\right) ,\qquad c_s = X_s-\frac{w}{1+z+w} ,\qquad  c_{c,b,t} =X_{c,b,t} .
\end{equation}
Let $c_i$, $i=\{1,\dots,5\}$, refer to the isoscalar couplings $\{u+d,s,c,b,t\}$, then one finds
\begin{equation}\label{eq:amusu2}
a_\mu = c_{u-d} \frac{\partial_\mu a}{2f_a} \tau_3,\quad a_{\mu,i}^{(s)}  = c_i \frac{\partial_\mu a}{2f_a} \mathbbm{1} \qquad\qquad \text{in SU(2)}.
\end{equation}
In any of the following equations, a summation over repeated $i$ is implied. Accordingly, the flavor subspace separation, this time with respect to the three lightest quarks, yields for the SU(3) case 
\begin{align}
\begin{split}
  \mathcal{L}_{aq} = - \left(\bar{q}_L\mathcal{M}_a q_R + \text{h.c.}\right) & + \left(\bar{q}\gamma^\mu  \gamma_5 \frac{\partial_\mu a}{2f_a} \left(c^{(1)}\mathbbm{1}+c^{(3)} \lambda_3+c^{(8)}\lambda_8\right) q\right)_{q=(u,d,s)^\text{T}}\\ & +\sum_{q=\{c,b,t\}}\left(\bar{q}\gamma^\mu  \gamma_5 \frac{\partial_\mu a}{2f_a} X_q q\right),
  \end{split}
\end{align}
where $\lambda_3$ and $\lambda_8$ are the third and eigth Gell-Mann matrices. Now
\begin{align}
\begin{split}
c^{(1)} & = \frac{1}{3}\left(X_u+X_d+X_s-1\right) , \qquad 
c^{(3)} = \frac{1}{2}\left(X_u-X_d-\frac{1-z}{1+z+w}\right), \\ 
c^{(8)} & = \frac{1}{2\sqrt{3}}\left(X_u+X_d-2X_s-\frac{1+z-2w}{1+z+w}\right)
  \end{split}
\end{align}
and 
\begin{equation}\label{eq:amusu3}
  a_\mu  = \frac{\partial_\mu a}{2f_a} \left(c^{(3)} \lambda_3+c^{(8)}\lambda_8\right) ,\quad
    a^{(s)}_{\mu,i} = c_i \frac{\partial_\mu a}{2f_a} \mathbbm{1} \qquad\qquad \text{in SU(3)},
\end{equation}
where this time $c_i=\{c^{(1)}, c_c, c_b, c_t\}$.
\section{Heavy baryon chiral perturbation theory with axions}
\subsection{Basic definitions}
The baryon fields $B$ and meson fields $\Phi$ are collected in
\begin{equation}
B  = N = \binom{p}{n},\qquad \Phi = \begin{pmatrix}\pi^0 & \sqrt{2}\pi^+\\ \sqrt{2}\pi^- & -\pi^0\end{pmatrix}
\end{equation}
in the case of SU(2) HBCHPT, whereas
\begin{equation}
B = \begin{pmatrix}
\frac{1}{\sqrt{2}} \Sigma_3  + \frac{1}{\sqrt{6}} \Lambda_8 & \Sigma^+ & p \\
\Sigma^- & -\frac{1}{\sqrt{2}} \Sigma_3 + \frac{1}{\sqrt{6}} \Lambda_8 & n \\
\Xi^- & \Xi^0 & -\frac{2}{\sqrt{6}} \Lambda_8 \end{pmatrix},\ \Phi = \begin{pmatrix}
\pi_3 + \frac{1}{\sqrt{3}} \eta_8 & \sqrt{2} \pi^+ & \sqrt{2} K^+ \\
\sqrt{2} \pi^- & - \pi_3 + \frac{1}{\sqrt{3}} \eta_8 & \sqrt{2} K^0 \\
\sqrt{2} K^- & \sqrt{2} \bar{K^0} & -\frac{2}{\sqrt{3}} \eta_8\end{pmatrix}
\end{equation}
in SU(3), where in any case $\Phi$ appears in the unitary matrix
\begin{equation}
u=\sqrt{U}=\exp\left(i\frac{\Phi}{2F_p}\right).
\end{equation}
Note that the physical $\Sigma^0$, $\Lambda$, $\pi^0$, and $\eta$ are mixed states parameterized by the mixing angle $\epsilon$ constrained from the relation
\begin{equation}
\tan{2\epsilon} = \frac{\tr{\lambda_3 \mathcal{M}_q}}{\tr{\lambda_8 \mathcal{M}_q}}~.
\end{equation}
Here, the heavy baryon limit is applied meaning that $\mathcal{L}_{\Phi B}$ contains an expansion in the inverse baryon mass $m_B$ ( $=$ nucleon mass in the SU(2) case). Furthermore, any Dirac bilinear is entirely expressed by means of the baryon four-velocity $v_\mu$ and the spin operator $S_\mu=\tfrac{i}{2} \gamma_5 \sigma_{\mu\nu} v^\nu$. This means that the fields $B$ are actually velocity dependent fields, which more correctly would be denoted $B_v$.

The axion enters HBCHPT in the following basic building blocks:
\begin{align}
\begin{split}
u_\mu & = i\left[ u^\dagger \partial_\mu u - u \partial_\mu u^\dagger - i u^\dagger a_\mu u - i u a_\mu u^\dagger \right], \qquad u_{\mu, i}  = 2a^{(s)}_{\mu, i}, \\
\Gamma_\mu & = \frac{1}{2}\left[ u^\dagger \partial_\mu u + u \partial_\mu u^\dagger - i u^\dagger  a_\mu u + i u a_\mu u^\dagger \right],\\
\chi_\pm & = u^\dagger\chi u^\dagger \pm u\chi^\dagger u,\qquad\text{with}\ \chi=2B_0 \mathcal{M}_a.
\end{split}
\end{align}
These objects are $2\times 2$ matrices in the SU(2) case and  $3\times 3$ matrices in SU(3). The chiral connection is needed for the chiral covariant derivative
\begin{align}
\mathcal{D}_\mu N &  = \partial_\mu N + \Gamma_\mu N  & & \text{in SU(2)}, \\
\com{\mathcal{D}_\mu}{B} & = \partial_\mu B + \com{\Gamma_\mu}{B} & & \text{in SU(3)}.
\end{align}
\subsection{General form of the axion-baryon coupling}
As chiral Lagrangians are organized with respect to chiral orders
\begin{equation}\label{eq:Lexpansion}
  \mathcal{L}_{\Phi B} = \mathcal{L}_{\Phi B}^{(1)} + \mathcal{L}_{\Phi B}^{(2)} + \mathcal{L}_{\Phi B}^{(3)}
  + \dots + \mathcal{L}_{\Phi}^{(2)}+ \mathcal{L}_{\Phi}^{(4)} + \dots
\end{equation}
it is clear from Eqs.~\eqref{eq:Ma}, \eqref{eq:amusu2}, and \eqref{eq:amusu3} that any axion-baryon coupling can be organized in terms of inverse powers of the expectedly large parameter $f_a$, because
\begin{align}
a_\mu,~a_{\mu,i}^{(s)} = \mathcal{O}(1/f_a), \qquad
\mathcal{M}_a =\exp\left({i\frac{a}{f_a}\mathcal{Q}_a}\right)\,\mathcal{M}_q = \mathcal{M}_q + i\frac{a}{f_a}\frac{1}{\tr{\mathcal{M}_q^{-1}}} + \mathcal{O}(1/f_a^2).
\end{align}
certainly only leading terms $\propto 1/f_a$ can contribute significantly, so the general axion-baryon coupling can be expressed as
\begin{equation}
\raisebox{-0.8cm}{\includegraphics[height=1.8cm]{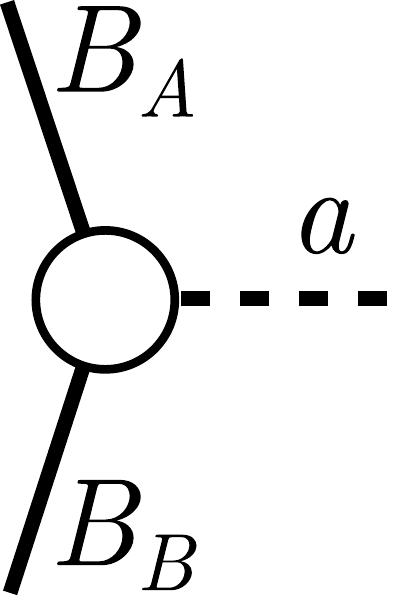}}\ = G_{aAB}\, \sdot{S}{q},\qquad\text{with}\ G_{aAB} = -\frac{1}{f_a} g_{aAB} + \mathcal{O}\left(\frac{1}{f_a^2}\right),
\end{equation}
where $A,B$ refer either to $p$ or $n$ (in SU(2)), or to the SU(3) indices of the octet baryons in the physical basis. Finally, $g_{aAB}$ contains an expansion in chiral power counting
\begin{equation}
 g_{aAB} = \underbrace{g^{(1)}_{aAB}}_\text{LO,tree} + \underbrace{g^{(2)}_{aAB}}_{\text{NLO},1/m_B}  + \underbrace{g_{aAB}^{(3)}}_{\text{NNLO},1/m_B^2,\text{one-loop}} + \dots
\end{equation}
\subsection{The case of SU(2) HBCHPT}
\begin{figure}
\centering
\includegraphics[width=0.47\textwidth]{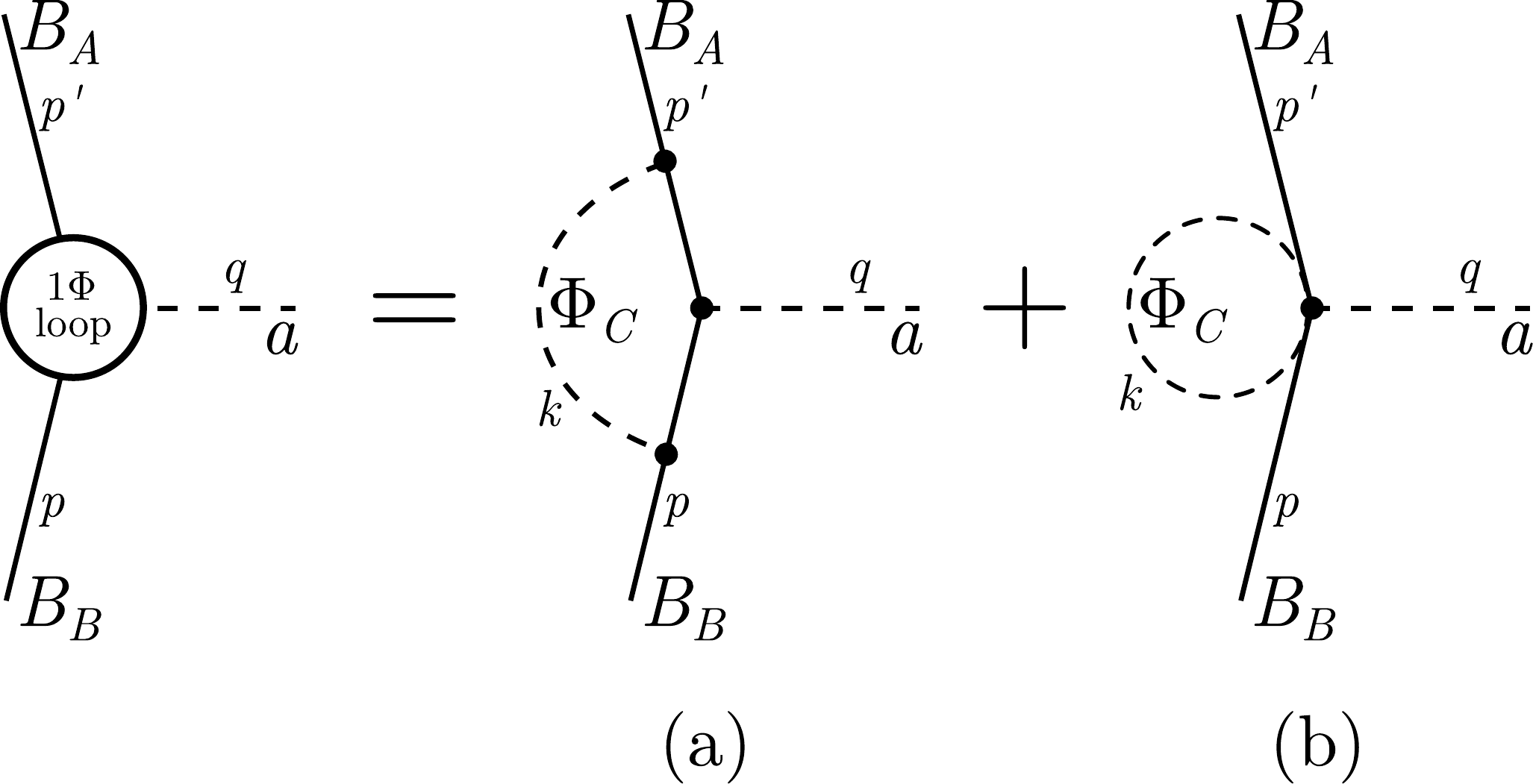}
\caption{Loop contributions to the axion-baryon coupling. In SU(2), $B_A$ and $B_B$ are either proton or neutron, while in SU(3) $A$, $B$ and $C$ are SU(3) indices in the physical basis for baryons and mesons.}
\label{fig:loops}
\end{figure}

The relevant pieces of the SU(2) HBCHPT Lagrangian up to next-to-next-to-leading order are
\begin{align}
\begin{split}
  & \mathcal{L}_{\pi N} = \bar{N} \biggl\{  i\sdot{v}{\mathcal{D}}+g_A \sdot{S}{u} + g_0^i \sdot{S}{u_i} + \mathcal{L}_{1/m_N} + \mathcal{L}_{1/m_N^2} \\ &
  \	+d_{16}(\lambda) \sdot{S}{u} \tr{\chi_+}+d_{16}^i(\lambda) \sdot{S}{u_i} \tr{\chi_+}+d_{17} S^\mu \tr{u_\mu\chi_+} +id_{18} S^\mu\com{\mathcal{D}_\mu}{\chi_{-}} \\ &
  \ + id_{19}S^\mu\com{\mathcal{D}_\mu}{\tr{\chi_-}}+\tilde{d}_{25}(\lambda)\sdot{v}{\overset{\leftarrow}{\mathcal{D}}} \sdot{S}{u}\sdot{v}{ \mathcal{D}}+\tilde{d}^i_{25}(\lambda)\sdot{v}{\overset{\leftarrow}{\mathcal{D}}}\sdot{S}{u_i} \sdot{v}{\mathcal{D}}\\
  & \ +\tilde{d}_{29}(\lambda)\left(S^\mu\left[\sdot{v}{\mathcal{D}}, u_\mu\right] \sdot{v}{\mathcal{D}} + \text{h.c.}\right) +\tilde{d}_{29}^i(\lambda)\left(S^\mu\left[\sdot{v}{\mathcal{D}}, u_{\mu,i}\right] \sdot{v}{\mathcal{D}} + \text{h.c.}\right) \biggr\} N ,
\end{split}
\end{align}
where $\mathcal{L}_{1/m_N}$ and $\mathcal{L}_{1/m_N^2}$ are the terms of the $1/m_N$ expansion, $g_A$ and the $g_0^i$'s are the axial isovector and isoscalar coupling constants, and the $d_j^{(i)}$ and $\tilde{d}_j^{(i)}$ are low-energy constants, from which some depend on the scale $\lambda$ and are needed for the renormalization of the one-pion loop contributions shown in Fig.~\ref{fig:loops}. The full renormalized NNLO result for the axion-nucleon coupling is (here given in nucleon rest frame, i.e. $v=(1,0,0,0)^\text{T}$  and $q_0=\sdot{v}{q} \ll m_N$)
\begin{align}
\begin{split}
g_{aNN} = & g_a\left(1+\frac{q_0}{2m_N} +\frac{q_0^2}{4m_N^2} \right) + \frac{\hat{g}_a}{6}\left(\frac{g_A M_\pi}{4\pi F_\pi}\right)^2\\  
&\quad\times \left[-1 +\left(\frac{q_0}{M_\pi}\right)^2 + \frac{2}{q_0 M_\pi^2}\left(\frac{\pi M_\pi^3}{2} -\left(M_\pi^2-q_0^2\right)^\frac{3}{2}\arccos\frac{q_0}{M_\pi}\right)\right]\\ 
& + 4M_\pi^2 \biggl[\left(\bar{d}_{16} \tau_3+d_{17}\frac{m_u-m_d}{m_u+m_d}\right) c_{u-d} + \bar{d}_{16}^i c_i-\left(d_{18}+2d_{19}\right)\frac{m_um_d}{(m_u+m_d)^2}\biggr], 
\end{split}
\end{align}
where the first term is the leading order result with the first two terms of the $1/m_N$ expansion. The $\bar{d}_j^{(i)}$ denote the renormalized and scale independent low-energy constants, and
\begin{equation}
g_a  = g_A c_{u-d} \tau_3 + g_0^i c_i \mathbbm{1}, \quad
\hat{g}_a  = - g_A c_{u-d} \tau_3 + 3 g_0^i c_i \mathbbm{1}\ .
\end{equation}
The couplings $g_A$ and the $g_0^i$'s can be matched to the nucleon matrix elements, i.\,e.
\begin{equation}
g_A = \Delta u-\Delta d,\qquad g_0^{u+d} = \Delta u+\Delta d, \qquad g_0^q = \Delta q~,\text{ for } q=s,c,b,t
\end{equation}
where $s^\mu\Delta q=\langle p | \bar{q}\gamma^\mu\gamma_5 q|p\rangle$, with $s^\mu$ the spin of the proton. The leading order results for the axion-proton and axion-neutron couplings then can be written as
\begin{align}\label{eq:SU2result}
\begin{split}
g_{app}^{(1)} & = -\frac{\Delta u+z\Delta d+w\Delta s}{1+z+w}+\Delta u X_u + \Delta d X_d+\sum_{q=\{s,c,b,t\}} \Delta q X_q \\
g_{ann}^{(1)} & = -\frac{z\Delta u+\Delta d+w\Delta s}{1+z+w}+\Delta d X_u + \Delta u X_d+\sum_{q=\{s,c,b,t\}} \Delta q X_q
\end{split}
\end{align} 
The numerical results are discussed below when it comes to the SU(3) results.
\subsection{The case of SU(3) HBCHPT}
In principle, a complete $O(p^3)$ description of the axion-baryon coupling would also include terms from the $1/m_B$ expansion, the NNLO Lagrangian and the topologically same diagrams as in Fig.~\ref{fig:loops}. Here, I restrict the discussion to the leading order calculation based on the Lagrangian (see Ref.~\cite{Vonk:2021sit} for more details)
\begin{equation}
  \mathcal{L}_{\Phi B}^{(1)} = \tr{ i \bar{B} v^\mu \com{\mathcal{D}_\mu}{B}} + D \tr{\bar{B} S^\mu \acom{u_\mu}{B}}
  + F\tr{\bar{B} S^\mu \com{u_\mu}{B}} + D^i \tr{\bar{B} S^\mu u_{\mu,i} B} ,
\end{equation}
which as in the SU(2) case comes with axial isovector ($F$, $D$) and isoscalar couplings ($D^i$). The leading order coupling of an axion to any baryon in the physical basis can be written as 
\begin{align}
\begin{split}
g_{aAB}^{(1)} = \frac{1}{2}\biggl\{ & D\left(c^{(3)} \tr{\lt{A}^\dagger \acom{\lambda_3}{\lt{B}}}+c^{(8)}
\tr{\lt{A}^\dagger \acom{\lambda_8}{\lt{B}}}\right)\\
& + F\left(c^{(3)} \tr{\lt{A}^\dagger \com{\lambda_3}{\lt{B}}}+c^{(8)} \tr{\lt{A}^\dagger
\com{\lambda_8}{\lt{B}}}\right) + 2 c_i D^i \delta_{AB}\biggr\},
\end{split}
\end{align}
where $\lt{A}$ are a set of traceless, non-Hermitian matrices, which are the generators of the physical basis. The coupling constants $F$, $D$, and $D^i$ are again matched to the nucleon matrix elements
\begin{align}\label{eq:DFchoice}
(D+F) & = g_A = \Delta u-\Delta d, &
-(D-3F)  &= \Delta u+\Delta d-2\Delta s, \\
D^1 & = \Delta u+\Delta d+\Delta s, &
D^q &= \Delta q,\qquad\text{for } q=c,b,t \nonumber
\end{align}
With this matching one exactly reproduces Eq.~\eqref{eq:SU2result} for the axion-proton and axion-neutron coupling, which means that at leading order the SU(2) and SU(3) results for these couplings are identical. Inserting \cite{Aoki:2021kgd}
\begin{align}
\begin{split}
\Delta u & = 0.847(50)\qquad \Delta d  = -0.407(34) \qquad \Delta s  = -0.035(13) \\
z & = 0.485(19) \qquad w = 0.025(1)
\end{split}
\end{align}
one gets the numerical results
\begin{align}
\begin{split}
g^{(1)}_{a\Sigma^+\Sigma^+}	& = -0.543(34)+0.847(50) X_u -0.035(13) X_d -0.407(34) X_s \\
g^{(1)}_{a\Sigma^-\Sigma^-}	& = -0.242(21)-0.035(13) X_u +0.847(50) X_d -0.407(34) X_s \\
g^{(1)}_{a\Sigma^0\Sigma^0}	& = -0.396(25)+0.417(25) X_u +0.395(25 )X_d -0.407(35) X_s \\
g^{(1)}_{app} 				& = -0.430(36)+0.847(50) X_u -0.407(34) X_d -0.035(13) X_s \\
g^{(1)}_{a\Xi^-\Xi^-} 		& = \phantom{-}0.140(15) -0.035(13) X_u -0.407(34) X_d +0.847(50) X_s \\
g^{(1)}_{ann} 				& = -0.002(30)-0.407(34) X_u + 0.847(50)X_d -0.035(13) X_s\\
g^{(1)}_{a\Xi^0\Xi^0} 		& = \phantom{-}0.267(23) -0.407(34) X_u -0.035(13) X_d +0.847(50) X_s \\
g^{(1)}_{a\Lambda\Lambda} 	& = \phantom{-}0.126(25) -0.147(25) X_u -0.125(25) X_d +0.677(35) X_s  \\
g^{(1)}_{a\Sigma^0\Lambda} 	& = -0.153(10)+0.463(25) X_u -0.476(25) X_d + 0.013(1) X_s
\end{split}
\end{align}
with the model dependent axion-quark couplings $X_q$. In particular, the constant first terms in each expression represent the KSVZ axion-baryon couplings (where $X_q=0$), while for the DFSZ one finds
\begin{align}
\begin{split}
g^{(1),\text{DFSZ}}_{a\Sigma^+\Sigma^+}	& = - 0.690(36) + 0.430(21) \sin^2\beta~,~~
g^{(1),\text{DFSZ}}_{a\Sigma^-\Sigma^-}	 = - 0.095(29) - 0.158(21) \sin^2\beta\\
g^{(1),\text{DFSZ}}_{a\Sigma^0\Sigma^0}	& = - 0.400(29) + 0.143(12) \sin^2\beta~,~~
g^{(1),\text{DFSZ}}_{app} 				 = - 0.577(38) + 0.430(21) \sin^2\beta\\
g^{(1),\text{DFSZ}}_{a\Xi^-\Xi^-} 		& = \phantom{-} 0.287(25) - 0.158(21) \sin^2\beta~,~~
g^{(1),\text{DFSZ}}_{ann} 				 = \phantom{-} 0.269(34) - 0.406(21) \sin^2\beta\\
g^{(1),\text{DFSZ}}_{a\Xi^0\Xi^0} 		& = \phantom{-} 0.531(29) - 0.406(21) \sin^2\beta~,~~
g^{(1),\text{DFSZ}}_{a\Lambda\Lambda} 	 = \phantom{-} 0.310(29) - 0.233(12) \sin^2\beta\\
g^{(1),\text{DFSZ}}_{a\Sigma^0\Lambda} 	& = - 0.308(13) + 0.309(16) \sin^2\beta\, .
\end{split}
\end{align}
\section{Conclusion}
The numerical results at the end of the last section have some notable consequences. First, the coupling of the axion to the neutron can vanish (or might at least be strongly suppressed) in both models (in the DFSZ at $\sin^2\beta\approx2/3$), whereas the axion-proton coupling is non-vanishing in any case. Second, the coupling to hyperons are of similar strength as to nucleons, which suggests that it might be advisable to include interactions of axions with these particles also in studies dedicated to neutron star cooling due to  axion bremsstrahlung.

In this brief review, I have only shown the $\mathcal{O}(p^3)$ results including one-pion loops for the SU(2) case, while one-meson loops have also been studied for the SU(3) case (see Ref.~\cite{Vonk:2021sit}). However, the numerical results for these cases suffer from the fact that many of the involved low energy constants (especially but not exclusively the isoscalar ones) are entirely unknown. Future work might fill this gap, which would make a more precise determination of the axion-baryon couplings possible.

\begin{acknowledgments}
I thank Ulf-G.\,Mei\ss{}ner and Feng-Kun Guo for the great collaboration this contribution is based on. Moreover, I thank the organizers of the RDP Online Workshop ``Aspects of Symmetry'', especially Akaki Rusetsky and Mirian Tabidze, for the opportunity to present these research results. This work is supported in part by the Deutsche Forschungsgemeinschaft (DFG) and the National Natural Science Foundation of China (NSFC) through the funds provided to the Sino-German Collaborative Research Center ``Symmetries and the Emergence of Structure in QCD'' (NSFC Grant No.\,12070131001, DFG Project-ID 196253076 -- TRR 110).
\end{acknowledgments}

\bibliographystyle{JHEP}
\bibliography{axion-baryon_PoS.bib}

\providecommand{\href}[2]{#2}\begingroup\raggedright\begin{thebibliography}{10}

\bibitem{Belavin:1975fg}
A.A.~Belavin, A.M.~Polyakov, A.S.~Schwartz and Y.S.~Tyupkin,
  \emph{{Pseudoparticle Solutions of the Yang-Mills Equations}},
  \href{https://doi.org/10.1016/0370-2693(75)90163-X}{\emph{Phys. Lett. B}
  {\bfseries 59} (1975) 85}.

\bibitem{Callan:1976je}
C.G.~Callan, Jr., R.F.~Dashen and D.J.~Gross, \emph{{The Structure of the Gauge
  Theory Vacuum}},
  \href{https://doi.org/10.1016/0370-2693(76)90277-X}{\emph{Phys. Lett. B}
  {\bfseries 63} (1976) 334}.

\bibitem{Callan:1977gz}
C.G.~Callan, Jr., R.F.~Dashen and D.J.~Gross, \emph{{Toward a Theory of the
  Strong Interactions}},
  \href{https://doi.org/10.1103/PhysRevD.17.2717}{\emph{Phys. Rev. D}
  {\bfseries 17} (1978) 2717}.

\bibitem{tHooft:1976rip}
G.~'t~Hooft, \emph{{Symmetry Breaking Through Bell-Jackiw Anomalies}},
  \href{https://doi.org/10.1103/PhysRevLett.37.8}{\emph{Phys. Rev. Lett.}
  {\bfseries 37} (1976) 8}.

\bibitem{tHooft:1976snw}
G.~'t~Hooft, \emph{{Computation of the Quantum Effects Due to a
  Four-Dimensional Pseudoparticle}},
  \href{https://doi.org/10.1103/PhysRevD.14.3432}{\emph{Phys. Rev. D}
  {\bfseries 14} (1976) 3432}.

\bibitem{Baluni:1978rf}
V.~Baluni, \emph{{CP Violating Effects in QCD}},
  \href{https://doi.org/10.1103/PhysRevD.19.2227}{\emph{Phys. Rev. D}
  {\bfseries 19} (1979) 2227}.

\bibitem{Lee:2020tmi}
D.~Lee, U.-G.~Mei\ss{}ner, K.A.~Olive, M.~Shifman and T.~Vonk,
  \emph{{\ensuremath{\theta} -dependence of light nuclei and nucleosynthesis}},
  \href{https://doi.org/10.1103/PhysRevResearch.2.033392}{\emph{Phys. Rev.
  Res.} {\bfseries 2} (2020) 033392}
  [\href{https://arxiv.org/abs/2006.12321}{{\ttfamily 2006.12321}}].

\bibitem{Elhatisari:2021eyg}
S.~Elhatisari, T.A.~L\"ahde, D.~Lee, U.-G.~Mei\ss{}ner and T.~Vonk,
  \emph{{Alpha-alpha scattering in the Multiverse}},
  \href{https://arxiv.org/abs/2112.09409}{{\ttfamily 2112.09409}}.

\bibitem{Kim:2008hd}
J.E.~Kim and G.~Carosi, \emph{{Axions and the Strong CP Problem}},
  \href{https://doi.org/10.1103/RevModPhys.82.557}{\emph{Rev. Mod. Phys.}
  {\bfseries 82} (2010) 557} [\href{https://arxiv.org/abs/0807.3125}{{\ttfamily
  0807.3125}}].

\bibitem{Guo:2015tla}
F.-K.~Guo, R.~Horsley, U.-G.~Mei\ss{}ner, Y.~Nakamura, H.~Perlt, P.E.L.~Rakow
  et~al., \emph{{The electric dipole moment of the neutron from 2+1 flavor
  lattice QCD}},
  \href{https://doi.org/10.1103/PhysRevLett.115.062001}{\emph{Phys. Rev. Lett.}
  {\bfseries 115} (2015) 062001}
  [\href{https://arxiv.org/abs/1502.02295}{{\ttfamily 1502.02295}}].

\bibitem{nEDM:2020crw}
{\scshape nEDM} collaboration, \emph{{Measurement of the permanent electric
  dipole moment of the neutron}},
  \href{https://doi.org/10.1103/PhysRevLett.124.081803}{\emph{Phys. Rev. Lett.}
  {\bfseries 124} (2020) 081803}
  [\href{https://arxiv.org/abs/2001.11966}{{\ttfamily 2001.11966}}].

\bibitem{Ubaldi:2008nf}
L.~Ubaldi, \emph{{Effects of theta on the deuteron binding energy and the
  triple-alpha process}},
  \href{https://doi.org/10.1103/PhysRevD.81.025011}{\emph{Phys. Rev. D}
  {\bfseries 81} (2010) 025011}
  [\href{https://arxiv.org/abs/0811.1599}{{\ttfamily 0811.1599}}].

\bibitem{Alexandrou:2020bkd}
C.~Alexandrou, J.~Finkenrath, L.~Funcke, K.~Jansen, B.~Kostrzewa, F.~Pittler
  et~al., \emph{{Ruling Out the Massless Up-Quark Solution to the Strong
  $\pmb{CP}$ Problem by Computing the Topological Mass Contribution with
  Lattice QCD}},
  \href{https://doi.org/10.1103/PhysRevLett.125.232001}{\emph{Phys. Rev. Lett.}
  {\bfseries 125} (2020) 232001}
  [\href{https://arxiv.org/abs/2002.07802}{{\ttfamily 2002.07802}}].

\bibitem{Peccei:1977hh}
R.D.~Peccei and H.R.~Quinn, \emph{{CP Conservation in the Presence of
  Instantons}}, \href{https://doi.org/10.1103/PhysRevLett.38.1440}{\emph{Phys.
  Rev. Lett.} {\bfseries 38} (1977) 1440}.

\bibitem{Peccei:1977ur}
R.D.~Peccei and H.R.~Quinn, \emph{{Constraints Imposed by CP Conservation in
  the Presence of Instantons}},
  \href{https://doi.org/10.1103/PhysRevD.16.1791}{\emph{Phys. Rev. D}
  {\bfseries 16} (1977) 1791}.

\bibitem{Weinberg:1977ma}
S.~Weinberg, \emph{{A New Light Boson?}},
  \href{https://doi.org/10.1103/PhysRevLett.40.223}{\emph{Phys. Rev. Lett.}
  {\bfseries 40} (1978) 223}.

\bibitem{Wilczek:1977pj}
F.~Wilczek, \emph{{Problem of Strong $P$ and $T$ Invariance in the Presence of
  Instantons}}, \href{https://doi.org/10.1103/PhysRevLett.40.279}{\emph{Phys.
  Rev. Lett.} {\bfseries 40} (1978) 279}.

\bibitem{Preskill:1982cy}
J.~Preskill, M.B.~Wise and F.~Wilczek, \emph{{Cosmology of the Invisible
  Axion}}, \href{https://doi.org/10.1016/0370-2693(83)90637-8}{\emph{Phys.
  Lett. B} {\bfseries 120} (1983) 127}.

\bibitem{Abbott:1982af}
L.F.~Abbott and P.~Sikivie, \emph{{A Cosmological Bound on the Invisible
  Axion}}, \href{https://doi.org/10.1016/0370-2693(83)90638-X}{\emph{Phys.
  Lett. B} {\bfseries 120} (1983) 133}.

\bibitem{Dine:1982ah}
M.~Dine and W.~Fischler, \emph{{The Not So Harmless Axion}},
  \href{https://doi.org/10.1016/0370-2693(83)90639-1}{\emph{Phys. Lett. B}
  {\bfseries 120} (1983) 137}.

\bibitem{Ipser:1983mw}
J.~Ipser and P.~Sikivie, \emph{{Are Galactic Halos Made of Axions?}},
  \href{https://doi.org/10.1103/PhysRevLett.50.925}{\emph{Phys. Rev. Lett.}
  {\bfseries 50} (1983) 925}.

\bibitem{Turner:1986tb}
M.S.~Turner, \emph{{Thermal Production of Not SO Invisible Axions in the Early
  Universe}}, \href{https://doi.org/10.1103/PhysRevLett.59.2489}{\emph{Phys.
  Rev. Lett.} {\bfseries 59} (1987) 2489}.

\bibitem{Duffy:2009ig}
L.D.~Duffy and K.~van Bibber, \emph{{Axions as Dark Matter Particles}},
  \href{https://doi.org/10.1088/1367-2630/11/10/105008}{\emph{New J. Phys.}
  {\bfseries 11} (2009) 105008}
  [\href{https://arxiv.org/abs/0904.3346}{{\ttfamily 0904.3346}}].

\bibitem{Marsh:2015xka}
D.J.E.~Marsh, \emph{{Axion Cosmology}},
  \href{https://doi.org/10.1016/j.physrep.2016.06.005}{\emph{Phys. Rept.}
  {\bfseries 643} (2016) 1} [\href{https://arxiv.org/abs/1510.07633}{{\ttfamily
  1510.07633}}].

\bibitem{Kim:1979if}
J.E.~Kim, \emph{{Weak Interaction Singlet and Strong CP Invariance}},
  \href{https://doi.org/10.1103/PhysRevLett.43.103}{\emph{Phys. Rev. Lett.}
  {\bfseries 43} (1979) 103}.

\bibitem{Shifman:1979if}
M.A.~Shifman, A.I.~Vainshtein and V.I.~Zakharov, \emph{{Can Confinement Ensure
  Natural CP Invariance of Strong Interactions?}},
  \href{https://doi.org/10.1016/0550-3213(80)90209-6}{\emph{Nucl. Phys. B}
  {\bfseries 166} (1980) 493}.

\bibitem{Dine:1981rt}
M.~Dine, W.~Fischler and M.~Srednicki, \emph{{A Simple Solution to the Strong
  CP Problem with a Harmless Axion}},
  \href{https://doi.org/10.1016/0370-2693(81)90590-6}{\emph{Phys. Lett. B}
  {\bfseries 104} (1981) 199}.

\bibitem{Zhitnitsky:1980tq}
A.R.~Zhitnitsky, \emph{{On Possible Suppression of the Axion Hadron
  Interactions. (In Russian)}}, {\emph{Sov. J. Nucl. Phys.} {\bfseries 31}
  (1980) 260}.

\bibitem{Donnelly:1978ty}
T.W.~Donnelly, S.J.~Freedman, R.S.~Lytel, R.D.~Peccei and M.~Schwartz,
  \emph{{Do Axions Exist?}},
  \href{https://doi.org/10.1103/PhysRevD.18.1607}{\emph{Phys. Rev. D}
  {\bfseries 18} (1978) 1607}.

\bibitem{Kaplan:1985dv}
D.B.~Kaplan, \emph{{Opening the Axion Window}},
  \href{https://doi.org/10.1016/0550-3213(85)90319-0}{\emph{Nucl. Phys. B}
  {\bfseries 260} (1985) 215}.

\bibitem{Srednicki:1985xd}
M.~Srednicki, \emph{{Axion Couplings to Matter. 1. CP Conserving Parts}},
  \href{https://doi.org/10.1016/0550-3213(85)90054-9}{\emph{Nucl. Phys. B}
  {\bfseries 260} (1985) 689}.

\bibitem{Georgi:1986df}
H.~Georgi, D.B.~Kaplan and L.~Randall, \emph{{Manifesting the Invisible Axion
  at Low-energies}},
  \href{https://doi.org/10.1016/0370-2693(86)90688-X}{\emph{Phys. Lett. B}
  {\bfseries 169} (1986) 73}.

\bibitem{GrillidiCortona:2015jxo}
G.~Grilli~di Cortona, E.~Hardy, J.~Pardo~Vega and G.~Villadoro, \emph{{The QCD
  axion, precisely}},
  \href{https://doi.org/10.1007/JHEP01(2016)034}{\emph{JHEP} {\bfseries 01}
  (2016) 034} [\href{https://arxiv.org/abs/1511.02867}{{\ttfamily
  1511.02867}}].

\bibitem{Vonk:2020zfh}
T.~Vonk, F.-K.~Guo and U.-G.~Mei\ss{}ner, \emph{{Precision calculation of the
  axion-nucleon coupling in chiral perturbation theory}},
  \href{https://doi.org/10.1007/JHEP03(2020)138}{\emph{JHEP} {\bfseries 03}
  (2020) 138} [\href{https://arxiv.org/abs/2001.05327}{{\ttfamily
  2001.05327}}].

\bibitem{Vonk:2021sit}
T.~Vonk, F.-K.~Guo and U.-G.~Mei\ss{}ner, \emph{{The axion-baryon coupling in
  SU(3) heavy baryon chiral perturbation theory}},
  \href{https://doi.org/10.1007/JHEP08(2021)024}{\emph{JHEP} {\bfseries 08}
  (2021) 024} [\href{https://arxiv.org/abs/2104.10413}{{\ttfamily
  2104.10413}}].

\bibitem{Kim:1986ax}
J.E.~Kim, \emph{{Light Pseudoscalars, Particle Physics and Cosmology}},
  \href{https://doi.org/10.1016/0370-1573(87)90017-2}{\emph{Phys. Rept.}
  {\bfseries 150} (1987) 1}.

\bibitem{Turner:1989vc}
M.S.~Turner, \emph{{Windows on the Axion}},
  \href{https://doi.org/10.1016/0370-1573(90)90172-X}{\emph{Phys. Rept.}
  {\bfseries 197} (1990) 67}.

\bibitem{Raffelt:1990yz}
G.G.~Raffelt, \emph{{Astrophysical methods to constrain axions and other novel
  particle phenomena}},
  \href{https://doi.org/10.1016/0370-1573(90)90054-6}{\emph{Phys. Rept.}
  {\bfseries 198} (1990) 1}.

\bibitem{DiLuzio:2020wdo}
L.~Di~Luzio, M.~Giannotti, E.~Nardi and L.~Visinelli, \emph{{The landscape of
  QCD axion models}},
  \href{https://doi.org/10.1016/j.physrep.2020.06.002}{\emph{Phys. Rept.}
  {\bfseries 870} (2020) 1} [\href{https://arxiv.org/abs/2003.01100}{{\ttfamily
  2003.01100}}].

\bibitem{Carenza:2020cis}
P.~Carenza, B.~Fore, M.~Giannotti, A.~Mirizzi and S.~Reddy, \emph{{Enhanced
  Supernova Axion Emission and its Implications}},
  \href{https://doi.org/10.1103/PhysRevLett.126.071102}{\emph{Phys. Rev. Lett.}
  {\bfseries 126} (2021) 071102}
  [\href{https://arxiv.org/abs/2010.02943}{{\ttfamily 2010.02943}}].

\bibitem{Fischer:2021jfm}
T.~Fischer, P.~Carenza, B.~Fore, M.~Giannotti, A.~Mirizzi and S.~Reddy,
  \emph{{Observable signatures of enhanced axion emission from protoneutron
  stars}}, \href{https://doi.org/10.1103/PhysRevD.104.103012}{\emph{Phys. Rev.
  D} {\bfseries 104} (2021) 103012}
  [\href{https://arxiv.org/abs/2108.13726}{{\ttfamily 2108.13726}}].

\bibitem{Tolos:2020aln}
L.~Tolos and L.~Fabbietti, \emph{{Strangeness in Nuclei and Neutron Stars}},
  \href{https://doi.org/10.1016/j.ppnp.2020.103770}{\emph{Prog. Part. Nucl.
  Phys.} {\bfseries 112} (2020) 103770}
  [\href{https://arxiv.org/abs/2002.09223}{{\ttfamily 2002.09223}}].

\bibitem{Aoki:2021kgd}
Y.~Aoki et~al., \emph{{FLAG Review 2021}},
  \href{https://arxiv.org/abs/2111.09849}{{\ttfamily 2111.09849}}.

\end{thebibliography}\endgroup
\end{document}